\documentclass[aps,prl, showpacs,twocolumn,twoside,floatfix,a4]{revtex4}
\usepackage{latexsym}
\usepackage{amsmath}
\usepackage{amssymb}
\usepackage{amsfonts}
\usepackage{comment}
\usepackage{bbm}
\usepackage{color}
\usepackage{epsfig}
\usepackage{pifont} 

\usepackage{graphicx,epsfig}

\newcommand{\<}{\langle}
\renewcommand{\>}{\rangle}

\begin{document}

\title{Quantum entanglement between the electron clouds of nucleic acids in DNA}

\author{Elisabeth Rieper$^1$\footnote{elisabeth.rieper@quantumlah.org},  Janet Anders$^{2}$ and Vlatko Vedral$^{1,3,4}$}

\address{$^1$ Center for Quantum Technologies, National University of Singapore, Republic of Singapore}
\address{$^2$ Department of Physics and Astronomy, University College London, London WC1E 6BT, United Kingdom}
\address{$^3$ Atomic and Laser Physics, Clarendon Laboratory, University of Oxford, Parks Road, Oxford OX13PU, United Kingdom}
\address{$^4$ Department of Physics, National University of Singapore, Republic of Singapore}

\date{\today}

\begin{abstract}

We model the electron clouds of nucleic acids in DNA as a chain of coupled quantum harmonic oscillators with dipole-dipole interaction between nearest neighbours resulting in a van der Waals type bonding. Crucial parameters in our model are the distances between the acids and the coupling between them, which we estimate from numerical simulations \cite{Cerny08}. We show that for realistic parameters nearest neighbour entanglement is present even at room temperature. We quantify the amount of entanglement in terms of negativity and single base von Neumann entropy. We find that the strength of the single base von Neumann entropy depends on the neighbouring sites, thus questioning the notion of treating single bases as logically independent units. We derive an analytical expression for the binding energy of the coupled chain in terms of entanglement and show the connection between entanglement and correlation energy, a quantity commonly used in quantum chemistry.
\end{abstract}


\maketitle

\section{Introduction}
The precise value of energy levels is of crucial importance for any kind of interaction in physics. This is also true for processes in biological systems.  It has recently been shown for the photosynthesis complex FMO~\cite{PhotosynExp, mohseni:174106, caruso:105106, Fassioli:2009uq} that maximum transport efficiency can only be achieved when the environment broadens the systems energy levels. Also for the olfactory sense the energy spectra of key molecules seem to have a more significant contribution than their shape \cite{Turin01121996}.  In  \cite{BP09} the possibility of intramolecular refrigeration is discussed.  A common theme of these works is the system's ability to use non-trivial quantum effects to optimise its energy levels.  This leads to the question whether a molecule's energy levels are only determined by its own structure, or if the environment {\it shapes} the molecule's energy level? Entanglement between system and environment is a necessary condition to alter the system's state. Here we study the influence of weak chemical bonds, such as intramolecular van der Waals interactions, on the energy level structure of DNA and discuss its connection to entanglement. To describe the van der Waals forces between the nucleic acids in a single strand of DNA, we consider a chain of coupled quantum harmonic oscillators.  Much work has been done investigating classical harmonic oscillators. However, this cannot explain quantum features of non-local interactions. Also, classical systems can absorb energy quanta at any frequency, whereas quantum systems are restricted to absorb energy quanta matching their own energy levels. This is of importance for site specific DNA-Protein interaction, as the probability of a protein to bind to a specific sequence of sites in DNA is governed by the relative binding energy \cite{Stormo:2010fk}.

Our work was motivated by a numerical study on the importance of dispersion energies in DNA \cite{Cerny08}. Dispersion energies describe attractive van der Waals forces between non-permanent dipoles. Recently their importance to stabilise macromolecules was realised \cite{Cerny,stacking}. 
Modelling macromolecules, such as DNA, is a tedious and complex task. It is currently nearly impossible to fully quantum mechanically simulate the DNA. Quantum chemistry has developed several techniques that allow the simulation of DNA with simplified dynamics. 
In \cite{Cerny08} the authors first quantum mechanically  optimise a small fragment of DNA in the water environment. Secondly, they {\it "performed various molecular dynamics (MD) simulations in explicit water based either fully on the empirical potential or on more accurate  QM/ MM MD simulations.
The molecular dynamics simulations were performed with an AMBER parm9916 empirical force field and the following modifications were introduced in the non-bonded part, which describes the potential energy of the system (see eq 1) and is divided into the electrostatic and Lennard-Jones terms. The former term is modelled by the Coulomb interaction of atomic point-charges, whereas the latter describes repulsion and dispersion energies,"}
\begin{equation}
V(r)=\frac{q_i q_j}{4 \pi \epsilon_0 r_{ij}}+4 \epsilon \left[ \left(\frac{\sigma}{r_{ij}} \right)^{12}- \left(\frac{\sigma}{r_{ij}} \right)^6 \right] \hbox{ ,}
\end{equation}
where the strength of the dispersion energy is scaled with the parameter $\epsilon$.
For $\epsilon=1$ the dynamics of the DNA strand is normal. For a weaker dispersion, $\epsilon=0.01$, there is in increase of $27\%$ in energy in the DNA. This increase of energy induces the unravelling of the double helix to a flat, ladder-like DNA. Many factors contribute to the spatial geometry of DNA, e.g. water interaction, the phosphate backbone, etc. However, one of the strongest contributions is the energy of the electronic degree of freedom inside a DNA strand, which is well shielded from interactions with water. Stronger interaction ($\epsilon=1$) allows the electrons clouds to achieve spatial configurations that require less structural energy. This allows a denser packing of the electron charges inside the double helix.
 \\ \\
Here we investigate with a simple model of DNA whether continuos variable entanglement can be present at room temperature, and how this entanglement is connected to the energy of the molecule. There are many technically advanced quantum chemically calculations for van der Waals type interaction, i.e.  \cite{PhysRevLett.91.233202}. The aim of this work is not to provide an accurate model, but to understand underlying quantum mechanical features and their role in this biological system. Also, there are many parallel developments between quantum information and quantum chemistry. This work bridges the concepts of entanglement and dispersion energies between the two fields. Finally, the advantages of quantifying chemical bonds in terms of entanglement were already mentioned in \cite{IJCH:IJCH5680470109}. Here we give the first example of a system whose chemical bonds are described by entanglement.

\section{Dispersion energies between nucleic acids}
\begin{figure}[t]
	\begin{center}
	\includegraphics[width=0.5\textwidth]{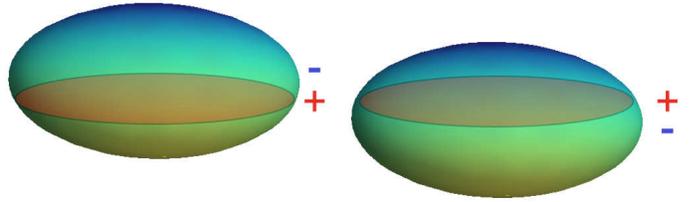}
\caption{\label{fig:scetchho} This graphic shows a sketch of a DNA nucleic acid. The mostly planar molecules are divided into the positively charged molecule core (red) and the negatively charged outer $\pi$ electron cloud (blue-yellow). In equilibrium the centre of both parts coincide, thus there is no permanent dipole. If the electron cloud oscillates around the core, a non permanent dipole is created \cite{DrudeModel}. The deviation out of equilibrium is denoted by $(x,y,z)$. The corresponding dipole is $\vec{\mu}=Q (x,y,z)$. This oscillation might be caused by an external field, or induced by quantum fluctuations, as it is given in a DNA strand.}
	\end{center}
\end{figure}

\begin{figure}[t]
	\begin{center}
	\includegraphics[width=0.5\textwidth]{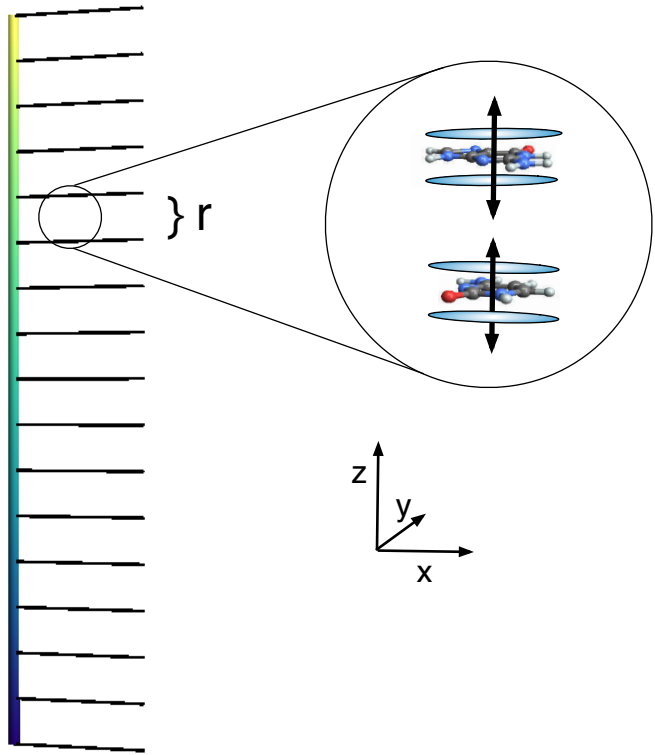}
	\caption{\label{fig:scetch} This graphic shows a sketch of a single DNA strand. The chain is along $z$ direction. Each bar in the single strand DNA  represents one nucleic acid: adenine, thymine, guanine or cytosine.  Around the core of atoms is the blue outer electron cloud. The oscillation of these electron clouds is modelled here as non-permanent harmonic dipoles, depicted by the arrows, with trapping potential $\Omega_{d}$ in dimension $d=x,y,z$.}
	\end{center}
\end{figure}

 The nucleic bases adenine, guanine, cytosine and thymine are planar molecules surrounded by $\pi$ electron clouds. We model each base as an immobile positively charged centre while the electron cloud is free to move around its equilibrium position, see Fig.~\ref{fig:scetchho}. There is no permanent dipole moment, while any displacement of the electron cloud creates a non-permanent dipole moment. Denoting the displacement of two centres by $(x,y,z)$, we  assume the deviation out of equilibrium $|(x,y,z)|$ to be small compared to the distance $r$ between neighboring bases in chain. The displacement of each electron cloud is approximated to second order and described by a harmonic oscillator with trapping potential $\Omega$ that quantifies the Coulomb attraction of the cloud to the positively charged centre. A single DNA strand resembles a chain of harmonic oscillators, see Fig.~ \ref{fig:scetch}, where each two neighboring bases with distance $r$ have dipole-dipole interaction.

The Hamiltonian for the DNA strand of N bases if given by
\begin{equation}
H=\sum_{j,d=x,y,z}^N \left(\frac{p_{j,d}^2}{2m}+\frac{m \Omega_d^2}{2}d_j^2+V_{j,dip-dip} \right)
\end{equation}
where $d$ denotes the dimensional degree of freedom, and the dipole potential 
\begin{equation}
V_{j, dip-dip}=\sqrt{\epsilon} \frac{1}{4 \pi \epsilon_0 r^3} \left( 3 (\vec{\mu}_j \cdot \vec{r}_N) (\vec{\mu}_{j+1} \cdot\vec{r}_N)-\vec{\mu}_j \cdot \vec{\mu}_{j+1}  \right)
\end{equation}
with $\vec{\mu}_j=Q (x_j, y_j, z_j)$ dipole vector of of site $j$ and $\vec{r}_N$ normalised distance vector between site $j$ and $j+1$. Due to symmetry $\vec{r}_N$ is independent  of $j$. The dimensionless scaling factor $\epsilon$ is varied to study the effects on entanglement and energy identical as in \cite{Cerny08}. In order to compare our model with \cite{Cerny08}, we consider 'normal' interaction, where the dipole-dipole interaction has full strength modelled by $\epsilon=1$ and 'scaled' interaction, where the dipole-dipole interaction is reduced to a hundredth of the original strength modelled by $\epsilon=0.01$. The distance between neighbouring bases in DNA is approximately $r_0=4.5 \AA$. For generality we will not fix the distance.\\
In general the single strand of DNA will not be perfectly linear and thus the dipole potential has coupling terms of the form $xz$ etc. Detailed analysis following \cite{1367-2630-12-2-025017} shows that
the energy contribution from the cross coupling terms is small, and they will be ignored here.
This leads to the interaction term 
\begin{equation}
V_{j, dip-dip}= \frac{Q^2}{4 \pi \epsilon_0 r^3} \left(  +x_j x_{j+1}    +y_j y_{j+1}  -2 z_j z_{j+1}  \right) \hbox{ .}
\end{equation}
The different signs for $x,y$ and $z$ reflect the orientation of the chain along $z$ direction. 

A discrete Fourier transformation of the form
\begin{eqnarray}
d_j = \frac{1}{\sqrt{N}}\sum_{l=1}^N e^{i\frac{2\pi}{N}jl}\tilde{d}_l\nonumber \\
p_{j,d} = \frac{1}{\sqrt{N}}\sum_{l=1}^N e^{-i\frac{2\pi}{N}jl}\tilde{p}_{l,d}
\end{eqnarray}
decouples the system into independent phonon modes. These modes can be diagonalized by introducing creation $a_{d,l}=\sqrt{\frac{m \Omega_d}{2 \hbar}} \left( \tilde{d}+\frac{i}{m \Omega_d} \tilde{p}_{l,d} \right)$ and annihilation operator $a^\dagger_{d,l}$.
This results in the dispersion relations
\begin{eqnarray}
\omega_{xl}^2 =\Omega_x^2 +2 \left(2\cos^2 \left(\frac{\pi l}{N}\right)-1\right)\frac{ Q^2}{4 \pi \epsilon_0 r^3m} \nonumber \\
\omega_{yl}^2 =\Omega_y^2 +2 \left(2\cos^2 \left(\frac{\pi l}{N}\right)-1\right)\frac{ Q^2}{4 \pi \epsilon_0 r^3m} \nonumber \\
\omega_{zl}^2 =\Omega_z^2 +4 \left(2\sin^2 \left(\frac{\pi l}{N}\right)-1\right)\frac{ Q^2}{4 \pi \epsilon_0 r^3m} 
\end{eqnarray}
and the Hamiltonian in diagonal form
\begin{equation}
H=\sum_{l=1,d=x,y,z}^N  \hbar \omega_{dl} \left(n_{d,l}+\frac{1}{2}\right)  \hbox{ ,}
\end{equation}
where $n_{d,l}=a^\dagger_{d,l}a_{d,l}$ is the number operator of mode $l$ in direction $d$. 

The trapping potentials $\Omega_d$ can be linked to experimental data (see table~\ref{data}) through the relation $\Omega_d=\sqrt{\frac{Q^2}{m_e \alpha_d}}$ , where $\alpha_d$ is the polarizability of the nucleid base. So far we did not discuss the number of electrons in the cloud. Both the trapping potential $\Omega_d^2$ as well as the interaction term $\frac{Q^2}{ m}$ depend linearly on the number of electrons, and thus the dispersion frequencies $\omega_{d,l}^2$ have the same dependance. The quantities of interest of this paper are entanglement and energy ratios, which are both given by ratios of different dispersion frequencies and are thus invariant of the number of electrons involved. In Table~\ref{data} we assumed the number of interacting electrons to be one, but our final results are independent of this special choice.

\begin{table}[htdp]
\caption{Numerical values for  polarizability of different nucleid acid bases \cite{HB89} in units of $1au=0.164\cdot 10^{-40}Fm^2$. The trapping frequencies are calculated using the formula $\Omega=\sqrt{\frac{Q^2}{m_e \alpha}}$ and are given in units of $10^{15}Hz$.   }
\label{tab:nucacid}
\begin{center}
\begin{tabular}{|c|c|c|c|c|c|c|}
\hline
nucleic acid & \phantom{.} $\alpha_x$  \phantom{.}& \phantom{.} $\alpha_y$ \phantom{.} & \phantom{.} $\alpha_z$  \phantom{.}& \phantom{x} $\Omega_x$ \phantom{x} &\phantom{x}$\Omega_y$\phantom{x} &\phantom{x}$\Omega_z$\phantom{x}\\
\hline 
adenine & 102.5 &114.0 & 49.6 & 4.1  &  3.9 & 6.0 \\
cytosine & 78.8   & 107.1 &44.2 & 4.7  &  4.1 & 6.3 \\
guanine & 108.7 &124.8  & 51.2 & 4.0 & 3.8  & 5.9\\
thymine & 80.7   & 101.7  & 45.9 & 4.7 & 4.2 & 6.2 \\
\hline
\end{tabular}
\end{center}
\label{data}
\end{table}%
Although the values for the four bases differ, all show similar $\Omega_x \approx \Omega_y$ (transverse), while there is an increase of $50\%$ in the longitudinal direction, $\Omega_z \approx \frac{3}{2} \Omega_{x,y}$. In the following we will approximate the chain to have the same value of  trapping potential at each base. In $x,y$ direction we will use $\Omega_{x,y}=4\cdot 10^{15}Hz$, and in $z$ direction $\Omega_{z}=6\cdot 10^{15}Hz$. 

\section{Entanglement and Energy} 
\begin{figure}[t]
	\begin{center}
	\includegraphics[width=0.5\textwidth]{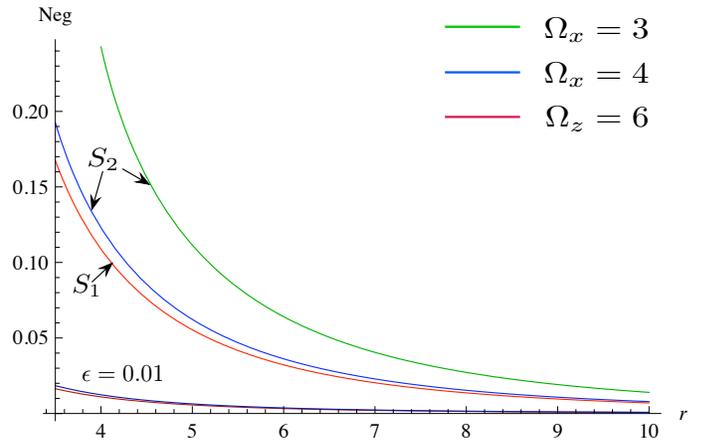}
	\caption{\label{fig:NegAll} This graphic shows the nearest neighbour negativity as a function of distance between sites in $\AA$ at $T=300K$. The three upper curves are for scaling factor $\epsilon=1$, the lower two curves are for scaling factor $\epsilon=0.01$. The red curve is for $z$ direction and $\Omega_z=6\cdot 10^{15}Hz$. The blue and green curve are for $x$ direction and $\Omega_x=4\cdot 10^{15}Hz$ and $\Omega_x=3\cdot 10^{15}Hz$.  The negativity for $\epsilon=0.01$ is much smaller than in the unscaled case. The amount of negativity strongly depends on the distance $r$ between sites and the value of trapping potential $\Omega$. The lower the potential, the higher the negativity. A typical distance between neighbouring base pairs in DNA is approximately $r=4.5 \AA$. Along the chain ($z$-direction) the $S_1$ criterion is violated, whereas transversal to the chain $S_2$ ($x$-direction) is violated. This reflects the geometry of the chain. Along the main axes of the chain energy is reduced by correlated movement. Transversal to the chain it is energetically better to be anti-correlated.}
	\end{center}
\end{figure}

We now clarify the influence of entanglement on energy. We will also derive an analytic expressions for the change in binding energy depending on entanglement witnesses.
 \\ \\
The chain of coupled harmonic oscillator is entangled at zero temperature, but is it possible to have entanglement at room temperature? There is a convenient way to calculate a criterion for nearest neighbour entanglement for harmonic chains \cite{PhysRevA.77.062102}, which compares the temperature $T$ with the coupling strength $\omega$ between neighbouring sites. In general, for $\frac{2 k_B T} {\hbar \omega}<1 $ one can expect entanglement to exist. Here the coupling between neighbouring clouds is given by $\omega=\sqrt{\sqrt{\epsilon} \frac{Q^2}{4 \pi \epsilon_0 m r^3}}\approx \epsilon^{1/4} 1.6\cdot 10^{15}Hz$ for $r=4.5 \AA$, which leads to $\frac{2 k_B 300K}{\hbar \omega}=0.05$ for $\epsilon=1$ and $0.16$ for $\epsilon=0.01$. This means that the coupling between electron clouds is dominant compared to the temperature, and thus implies the existence of entanglement even at biological temperatures.
An exact method to quantify the amount of entanglement in harmonic states it the violation of one of the two inequalities, related to the covariance matrix the state \cite{PhysRevA.70.022318}.
\begin{eqnarray}\label{entcrit}
0 \leq S_{1}=\frac{1}{\hbar^2}\left\langle (d_j + d_{j+1})^2\right\rangle
\left\langle (p_{d,j} - p_{d,j+1})^2\right\rangle-1 \\
0 \leq S_{2}=\frac{1}{\hbar^2}\left\langle (d_j - d_{j+1})^2\right\rangle
\left\langle (p_{d,j} +p_{d,j+1})^2\right\rangle-1
\end{eqnarray}
with $d_j$ position operator of site $j$ in direction $d$ and $p_{d,j}$ corresponding momentum operator. If one of the inequalities is violated, the sites $j$ and $j+1$ are entangled. The negativity, a widely used measure for entanglement, is calculated using the formula $Neg=\sum_{k=1}^2 \max\left[0,-\ln \sqrt{S_k+1}\right]$.  The negativity measures the amount of entanglement between two subsystems. It can be directly calculated from space and momentum operator expectation values, namely the above defined $S_{1,2}$ criteria.  The amount of negativity between neighbouring bases for room temperature is shown in Fig.~\ref{fig:NegAll}. For the normal coupling there is substantially more entanglement present than for the scaled interaction. This correlates with the amount of binding energy found in \cite{Cerny08}, where the DNA with normal coupling has a lower energy than the DNA with scaled coupling.

The above result motivates the question whether the binding energy can be expressed in terms of entanglement measures. In the limit of long distances, an analytical expression connects the amount of binding energy in the chain of oscillators with the values of $S_{1,2}$. Due to the strong coupling the chain of oscillators is effectively in its ground state, which we will assume in the following analysis.

The dispersion relations of the electron cloud oscillations can be expanded for large distances, i.e. $r^3 \rightarrow \infty$
\begin{equation}
\omega_{zl} \approx \Omega_z-4 \frac{ Q^2}{4 \pi \epsilon_0 m} \frac{1}{2 \Omega_z} \cos\left( \frac{ 2 \pi l}{N} \right) \frac{1}{r^3}+O\left[\frac{1}{r^6}\right]
\end{equation}
and similarly $1/\omega_{zl} $
\begin{equation}
\frac{1}{\omega_{zl}} \approx \frac{1}{\Omega_z}+4 \frac{ Q^2}{4 \pi \epsilon_0 m} \frac{1}{2 \Omega_z^3} \cos\left( \frac{ 2 \pi l}{N} \right) \frac{1}{r^3}+O\left[\frac{1}{r^6}\right]
\end{equation}
Inserting this expansion into the entanglement criterion $S_2$ gives:
\begin{equation}
S_{z,2} \approx -\frac{ Q^2}{ \pi \epsilon_0 m} \frac{1}{2 \Omega_z^2} \frac{1}{r^3} \hbox{ ,}
\end{equation}
while the corresponding expression for $S_{z,1}$ has a positive value. A similar expansion of the dispersion relation in $x$ direction leads to: 
\begin{equation}
S_{x,1} \approx -\frac{ Q^2}{2 \pi \epsilon_0 m} \frac{1}{2 \Omega_x^2} \frac{1}{r^3} \hbox{ .}
\end{equation}
This implies that nearest neighbor (n.n.) electronic clouds are entangled even at large distances. However the amount of entanglement decays very fast. 
We will now compare this result with the binding energy in the ground state. 
The binding energy is defined as the difference of energy of the entangled ground state and any hypothetical separable configuration
\begin{equation}
E_{z,bind}=\<\hat{H_z} \>-\sum_{I=1}^N \<\hat{H_z}_I \>=\hbar/2 \left( \sum_{l=1}^N \omega_{zl}- N \Omega_z \right) \hbox{ .}
\end{equation}
This definition is analogous to the definition of correlation energy in chemistry \cite{PhysicalChemistry}. The first approximation to the full Schr\"odinger equation is the Hartree-Fock equation and assumes that each electron moves independent of the others. Each of the electrons feels the presence of an average field made up by the other electrons. Then the electron orbitals are antisymmetrised. This mean field approach gives rise to a separable state, as antisymmetrisation  does not create entanglement. The Hartree-Fock energy is larger than the energy of the exact solution of the Schr\"odinger equation. The difference between the exact energy and the Hartree-Fock energy is called the correlation energy
\begin{equation}
E_{corr}=E_{exact}-E_{HF} \hbox{ .}
\end{equation}
Our definition of binding energy is a special case of the correlation energy, but we restrict our analysis here to phonons (bosons) instead of electrons. Our model describes the motional degree of freedom of electrons, namely the displacement of electron clouds out of equilibrium. 
We show for this special case that the amount of correlation energy is identical to entanglement measures.\\
Expanding the binding energy for $r^3 \rightarrow \infty$, the leading term is of order $\frac{1}{r^6}$ 
\begin{equation}\label{eq:Ez}
E_{z,bind} \approx \hbar/2 \left( - \left(\frac{ Q^2}{ \pi \epsilon_0 m}\right)^2 \frac{N}{16 \Omega_z^3} \frac{1}{r^6}\right)=-\frac{N \hbar \Omega_z}{8}S_2^2 \hbox{ ,}
\end{equation}
since the first order vanishes due to symmetry and similarly for $x$ direction:
\begin{equation}\label{eq:Ex}
E_{x,bind} \approx -\frac{N \hbar \Omega_x}{8}S_1^2 \hbox{ .}
\end{equation}
Eq.~\ref{eq:Ez}, \ref{eq:Ex} show a simple relation between the entanglement witnesses $S_{1,2}$ and the binding energy of the chain of coupled harmonic oscillators. The stronger the entanglement, the more binding energy the molecule has. Interestingly, along the chain the $S_1$ criterion is violated, whereas transversal to the chain $S_2$ is violated. This reflects the geometry of the chain. Along the main axes of the chain energy is reduced by correlated movement. Transversal to the chain it is energetically better to be anti-correlated. This means that the entanglement witnesses $S_{1,2}$ not only measure the amount of binding energy, but also the nature of correlation which gives rise to the energy reduction. This relation motivates the search for entanglement measures describing the binding energies of complex molecules. While the binding energy just measures energy differences the corresponding entanglement measures reflect more information. Without correlations between subsystems there would not be a chemical bond. It is precisely the purpose of entanglement measures not only to quantify, but also to characterise these correlations.

\section{Aperiodic potentials and information processing in DNA}
\begin{figure}[t]
	\begin{center}
	\includegraphics[width=0.5\textwidth]{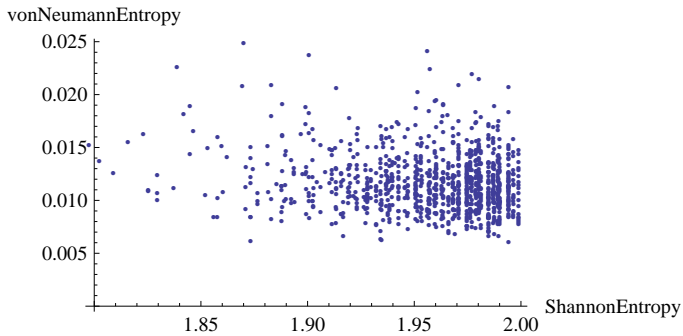}
	\caption{\label{fig:DistributionDNA} This graphic shows the average single site  von Neumann entropy of a chain of nucleic acids dependant on the classical Shannon entropy of the string.  Each string contains 50 bases with a random sequence of  A,C,G, or T. The distribution of nucleic acids determines the classical Shannon entropy. For each nucleic acid we used the value of polarizability of Table \ref{tab:nucacid} in x direction. The distance between sites is $r=4.5 \cdot 10^{-10} m$.  The plot has a sample size of 1000 strings. There is no direct correlation between quantum and classical information. The average amount of von Neumann entropy varies strongly for different sequences.}
	\end{center}
\end{figure}

In the above calculations we assumed a periodic potential, which allowed us to derive analytical solutions. Here we investigate the influence of
 aperiodic potentials and discuss the robustness of the previous conclusions. \\
Firstly we note that the potentials for different nucleic acids do not differ significantly, see table \ref{tab:nucacid}. Hence one would intuitively assume that a sequence of different local potentials changes the amount of entanglement but does not destroy it. To check this intuition more thoroughly, one can use the phonon frequencies of the aperiodic chain of oscillators. For a finite one-dimensional chain of 50 bases without periodic boundary conditions and with the sequence of nucleic acids randomly chosen, we solve the resulting coupling matrix numerically. The smallest dispersion frequency determines the thermal robustness; the smaller the frequencies $\omega_l$ the larger the probability that the thermal heat bath can excite the system. Sampling over 1000 randomly chosen sequences yielded $\min(\omega_{l})=3.2 \cdot 10^{15}Hz$ as smallest dispersion frequency. Comparing this with the thermal energy gives $\frac{2 k_B 300K}{\hbar \omega_l} \approx 0.03 $, which is still very small. \\
Thus the thermal energy is more than 20 times smaller than the smallest phonon frequency, which allows us to continue working with the ground state of the system.\\
Different sequences will cause fluctuations in the amount of entanglement in the chain of bases.
We determine for each string the average of single site von Neumann entropy and compare it with the classical amount of information measured by the Shannon entropy of each string.
The von Neumann entropy of a single site $j$  is obtained following \cite{PhysRevA.70.022318} with the formula
\begin{equation}
	S_V(r_j)=\frac{r_j+1}{2} \ln \left(\frac{r_j+1}{2}\right)-\frac{r_j-1}{2} \ln \left(\frac{r_j-1}{2} \right)
\end{equation}
where $r_j =\frac{1}{\hbar} \sqrt{\langle x_j^2 \rangle \langle p_{x_j}^2 \rangle }$, is the symplectic eigenvalue of the covariance matnrix of the reduced state. \\ 
To check whether the relative frequency of A,C,G and T influences the amount of entanglement within the coupled chain of oscillators, we also calculate the classical Shannon entropy of each string. Fig.~\ref{fig:DistributionDNA} shows the average amount of single site quantum entropy vs. classical entropy. There is, within this model, no direct correlation between classical and quantum entropy. For the same amount of Shannon entropy, i.e. same relative frequencies of A,C,G and T, the value of quantum correlations varies strongly between around $vNE=0.007$ and $vNE=0.025$. We note that for achieving a comparable amount of local disorder by thermal mixing a temperature of more than $2000 K$ is needed. This is a quantum effect without classical counterpart. Each base without coupling to neighbours would be in its ground state, as thermal energy is small compared to the energy spacing of the oscillators. As the coupling increases, the chain of bases evolves from a separable ground state to an entangled ground state. As a consequence of the global entanglement, each base becomes locally mixed. This feature cannot be reproduced by a classical description of vibrations. When a classical system is globally in the ground state, also each individual unit is in its ground state. Although it is already well known that globally entangled states are locally mixed, little is known about possible consequences for biological systems. In the following paragraph we discuss one such quantum effect on the information flow in DNA.\\ \\

How much information about the neighbouring sites is contained in the quantum degree of freedom of a single base? Is it accurate to describe a single nucleic acid as an individual unit or do the quantum correlations between bases require a combined approach of sequences of nucleic acids? The single site von Neumann entropy measures how strongly a single site is entangled with the rest of the chain and is therefore a good measure to answer this question. In the following we considered a string with 17 sites of a single strand DNA. Site 9 as well as sites 1-7 and 11-17 are taken to be Adenine. The identity of nucleic acids at sites 8 and 10 varies. Table \ref{tab:SingleSite} shows the resulting von Neumann entropy of site 9 dependent on its neighbours. The value of a single site depends on the direct neighbourhood. There is, for example, a distinct difference if an Adenine is surrounded by Cytosine and Thymine ($vNE=0.078$) or by Cytosine and Guanine ($vNE=0.084$). On the other hand, in this model there is little difference between Adenine and Guanine in site 8 and Guanine in site 10. Of course this model has not enough precision to realistically quantify how much a single site knows about its surroundings. Nevertheless it indicates that a single base should not be treated as an individual unit. When quantifying the information content and error channels of genetic information, the analysis is usually restricted to classical information transmitted through classical channels. While we agree that the genetic information is {\it stored} using classical information, e.g. represented by the set of molecules (A,C,G,T), we consider it more accurate to describe the {\it processing} of genetic information by quantum channels, as the interactions between molecules are determined by laws of quantum mechanics. 

\begin{table}[htdp]
\caption{Numerical values for  the von Neumann entropy of site 9 (Adenine) in a chain with open boundary condition containing 17 bases. The bases 1-7 and 11-17 are taken to be Adenine. The column gives the nucleic acid of site 8, the rows of site 10. The von Neumann entropy of site 9 varies with its neighbours. }
\label{tab:SingleSite}
\begin{center}
\begin{tabular}{|c|c|c|c|c|}
\hline
 & \phantom{.} Adenine\phantom{.} & \phantom{.}Cytosine  \phantom{.}& \phantom{.} Guanine \phantom{.} & \phantom{.}Thymine\\
\hline 
Adenine & 0.077& 0.082& 0.078& 0.081\\
Cytosine & 0.082 & 0.079 & 0.084& 0.078 \\
Guanine & 0.078& 0.084& 0.079 & 0.083\\
Thymine & 0.081 & 0.078 & 0.083& 0.078\\
\hline
\end{tabular}
\end{center}
\label{data}
\end{table}%

\section{Conclusions and discussion}
In this paper we modelled the electron clouds of nucleic acids in a single strand of DNA as a chain of coupled quantum harmonic oscillators with dipole-dipole interaction between nearest neighbours. Our main result is that the entanglement contained in the chain coincides with the binding energy of the molecule. We derived in the limit of long distances and periodic potentials analytic expressions linking the entanglement witnesses to the energy reduction due to the quantum entanglement in the electron clouds. Motivated by this result we propose to use entanglement measures to quantify correlation energy, a quantity commonly used in quantum chemistry. As the interaction energy given by $\hbar \omega$ is roughly 20 times larger than the thermal energy $k_B 300K$ the motional electronic degree of freedom is effectively in the ground state. Thus the entanglement persists even at room temperature. Additionally, we investigated the entanglement properties of aperiodic potentials. For randomly chosen sequences of A,C,G, or T we calculated the average von Neumann entropy. There exists no direct correlation between the classical information of the sequence and its average quantum information. The average amount of von Neumann entropy varies strongly, even among sequences having the same Shannon entropy. Finally we showed that a single base contains information about its neighbour, questioning the notion of treating individual DNA bases as independent bits of information.

\smallskip

\textit{Acknowledgments}: E.R. is supported by the National Research Foundation and Ministry of Education in Singapore.
 J.A. is supported by the Royal Society. V.V. acknowledges financial support from the Engineering and Physical Sciences Research Council, the Royal Society and the Wolfson Trust in UK as well as the National Research Foundation and Ministry of Education, in Singapore.

\smallskip
\smallskip
\bibliography{CVDNAcitelist.bib}

\end{document}